\documentclass[10pt, twocolumn]{IEEEtran}

\IfFileExists{multirow.sty}{\usepackage{multirow}}{}
\usepackage{booktabs}
\usepackage{longtable}

\IfFileExists{cite.sty}{\usepackage{cite}}{}

\usepackage{graphicx}
\usepackage{amsmath}
\usepackage{amsfonts}
\usepackage{amssymb}
\IfFileExists{stfloats.sty}{\usepackage{stfloats}}{}
\IfFileExists{arydshln.sty}{\usepackage{arydshln}}{}

\IfFileExists{subfigure.sty}{\usepackage{subfigure}}{}
\graphicspath{{figure/}}

\newtheorem{remark}{Remark}

\usepackage{amsfonts, amsmath, mathrsfs, amsbsy, amssymb, dsfont, color}
\usepackage{graphicx}

\newcommand\prefixtext[1]{%
  \ifvmode\else\\\@empty\fi
  \noalign{%
    \penalty0%
    \vbox{\mathstrut}%
    \penalty10000%
    \vskip-\baselineskip
    \penalty10000%
    \vbox to 0pt{%
      \normalbaselines
      \ifdim\linewidth=\columnwidth
      \else
        \parshape\@ne
        \@totalleftmargin\linewidth
      \fi
      \vss
      \noindent#1\par}%
      \penalty10000%
      \vskip-\baselineskip}%
      \penalty10000}

\newtheorem{lemma}{Lemma}
\newtheorem{proposition}{Proposition}

\newcommand{\qed}{\nobreak \ifvmode \relax \else
      \ifdim\lastskip<1.5em \hskip-\lastskip
      \hskip1.5em plus0em minus0.5em \fi \nobreak
      \vrule height0.75em width0.5em depth0.25em\fi}

\DeclareMathAlphabet{\mathpzc}{OT1}{pzc}{m}{it}
\def\({\left(}
\def\){\left)}

\def\bmu{\boldsymbol{\mu}}

\def\({\left(}
\def\){\right)}
\def\[{\left[}
\def\]{\right]}
\def\BEq{\begin{eqnarray}}
\def\EEq{\end{eqnarray}}
\def\BE*{\begin{eqnarray*}}
\def\EE*{\end{eqnarray*}}
\def\BA{\begin{array}}
\def\EA{\end{array}}

\def\Nn{\nonumber}
\def\0{\mathbf{0}}
\def\1{\mathbf{1}}

\def\G{\mathbf{G}}
\def\h{\mathbf{h}}

\def\and{\prefixtext{and}}

\def\Im{\mathrm{Im}}
\def\Re{\mathrm{Re}}

\title{One-Bit Coherent Distributed Radar Detection in Clutter via Velocity-Plane Search: Vanishing Loss and EEC-Calibrated Thresholding}

\author{Yu-Hang Xiao,~\IEEEmembership{Member,~IEEE}, David Ram{\'\i}rez,~\IEEEmembership{Senior Member,~IEEE},~Xiaochuan Ke, Lei Huang,~\IEEEmembership{Senior Member,~IEEE}, and Hing Cheung So,~\IEEEmembership{Fellow,~IEEE}
\thanks{\textcolor{blue}{This work has been submitted to the IEEE for possible publication. Copyright may be transferred without notice, after which this version may
no longer be accessible.}}
 \thanks{Y.-H. Xiao, Xiaochuan Ke and L. Huang  are with the State Key Laboratory of Radio Frequency Heterogeneous Integration, Shenzhen University, Shenzhen, 518060, China (e-mail: yuhangxiao@szu.edu.cn,xk2026@szu.edu.cn,lhuang@szu.edu.cn).}
 \thanks{D. Ram{\'\i}rez is with the Department of Signal Theory and Communications, Universidad Carlos III de Madrid, Madrid 28911, Spain, and also with the Gregorio Mara{\~n}{\'o}n Health Research Institute, Madrid 28007, Spain (email: david.ramirez@uc3m.es).}
 \thanks{H. C. So is with the Department of Electrical Engineering, City University of Hong Kong, Hong Kong, China (e-mail: hcso@ee.cityu.edu.hk).}
}

\begin{document}
\bstctlcite{paper_bib_control}

\IfFileExists{epsf.tex}{\input{epsf}}{}
\date{}

\maketitle

\begin{abstract}

Coherent distributed radar architectures enhance the detectability of low-reflectivity targets. However, conventional coherent processing generally entails prohibitively large volumes of inter-station data exchange, thereby motivating the adoption of one-bit quantization strategies. In this work, we investigate a clutter-dominated regime in which the interference is driven by a dense ensemble of weak scatterers. We demonstrate that, under these conditions, one-bit quantization can exhibit a \emph{vanishing performance loss} with respect to ideal high-resolution sampling in terms of detection capability, in contrast to the classical $2$~dB performance penalty associated with one-bit detection in white Gaussian noise environments.
For unknown target velocity, we reformulate joint velocity-search processing under the effective one-bit model. Maximizing over the velocity plane makes false-alarm calibration an extreme-value problem for a correlated two-dimensional random field, whose maximum is not characterized by its pointwise null distribution alone. We therefore use an expected Euler characteristic approximation to calibrate the search threshold. We obtain analytic expressions for a separable two-station geometry and use constrained coefficient fitting for larger networks. Monte Carlo simulations demonstrate negligible quantization loss in the clutter-dominated setting and conservative false-alarm calibration under the tested Gaussian slow-time model.
\end{abstract}

\begin{IEEEkeywords}
One-bit sampling, distributed radar coherent detection, velocity-plane search, expected Euler characteristic (EEC).
\end{IEEEkeywords}

\section{Introduction}

Widely separated radar stations observe a target from different views and exploit spatial diversity to mitigate radar cross-section (RCS) fluctuations~\cite{fishler2006spatial,haimovich2008mimo}. Centralized processing usually requires substantial receiver-data exchange. To reduce this communication burden, low-bit and mixed-bit schemes have been studied for distributed radar detection~\cite{xiang2023liquid,yang2023hybrid} and direct position determination~\cite{ni2024mixedbit}. Prior one-bit radar work has considered parameter estimation with time-varying thresholds~\cite{ameri2019onebit}, gridless angle--Doppler estimation~\cite{xi2020gridless}, joint waveform/filter design under uncertain interference statistics~\cite{bose2019waveform}, and target detection under white Gaussian noise~\cite{xiao2022onebit} and colored Gaussian noise~\cite{xiao2024colored}. Despite practical motivation, most existing one-bit radar studies focus on thermal-noise-dominated models and rarely account for clutter, even though clutter is often the dominant interference in many radar applications. In this paper, we consider one-bit coherent distributed detection in cluttered environments.

Cluttered environments are naturally more complicated than thermal-noise-only settings, making the effect of one-bit quantization on detection performance less straightforward. Under thermal-noise-dominated models, one-bit detection incurs a penalty of approximately $2$~dB in the low signal-to-noise ratio (SNR) regime~\cite{xiao2022onebit}. In the clutter-limited regime, however, the dominant interference is not thermal noise but the clutter return itself, which can be modeled as a dense superposition of many weak scatterers. When both the target echo and the pre-compression clutter echo are sufficiently weak at the analog-to-digital converter (ADC) input, that is, well below the thermal noise floor prior to range compression, one-bit sampling mainly introduces a common attenuation factor on these weak echoes, while the resulting quantization distortion can be treated as an additional noise-like term. Consequently, the relative strength between the target and clutter after coherent processing is essentially preserved, and the performance of the one-bit detector can approach that of an ideal high-resolution receiver. We formalize this intuition using the weak-input approximation and clutter dominance in the detection subspace, and show how the resulting loss approaches $0$~dB.

In a distributed radar, spatially separated transmitter--receiver paths observe different Doppler projections of a common target velocity~\cite{haimovich2008mimo}. Channel-wise Doppler estimation followed by velocity fitting, as in~\cite{xiang2023liquid}, may propagate local peak-selection errors into the fused velocity at low signal-to-clutter ratios (SCRs). Joint velocity-domain detection~\cite{he2010homogeneous} combines the path responses at the Dopplers implied by each candidate velocity before maximizing the aggregate statistic. This processing enforces geometric consistency during detection and retains evidence across stations without first committing to individual Doppler estimates. We reformulate this established approach under the effective slow-time model of the one-bit matched-filter outputs, obtaining a quadratic detection field over the two-dimensional velocity plane.

The resulting velocity search, however, introduces a challenge in false-alarm analysis: the false-alarm probability is no longer tied to a single velocity point, but to the tail behavior of the maximum of a correlated two-dimensional detection map. This correlation arises from applying similar Doppler templates to the same observations and must therefore be accounted for in false-alarm calibration. For related scan-and-maximize problems, Develter et al.~\cite{develter2023falsealarm} applied expected Euler characteristic (EEC) methods to continuous Doppler and Doppler--angle searches under white Gaussian noise, with a more comprehensive treatment including frequency-modulated continuous wave (FMCW) range--Doppler searches in Develter's dissertation~\cite[Ch.~3]{develter2023thesis}. They later used spherical tube geometry to analyze normalized matched-filter searches over a scalar target parameter~\cite{develter2024geometrical}. Here, we apply the EEC framework~\cite{adler2007randomfields,taylor2005validity} to the two-dimensional velocity field formed by summing station detection scores after one-bit processing. We derive analytic expressions for a separable two-station case and use constrained calibration for larger networks to extrapolate from moderate to high thresholds.

The contributions of this paper are summarized as follows:

\begin{enumerate}

\item \textbf{Vanishing quantization loss in clutter}: We show that one-bit detection can approach ideal high-resolution performance when target and clutter echoes are weak at the ADC input and clutter dominates the effective additive noise in the detection subspace after coherent processing. We establish this result for fixed-signature detection with known covariances under the effective Gaussian slow-time model, and verify negligible loss in velocity-search simulations.

\item \textbf{EEC-based threshold calibration for velocity-plane search}: Velocity-plane maximization makes false-alarm calibration depend on the upper tail of the maximum of a correlated two-dimensional random field. We develop an EEC-based approximation that accounts for the field's correlation structure and the search-domain geometry. We obtain analytic Euler characteristic (EC) densities for a separable two-station geometry and a constrained coefficient-calibration procedure for larger networks, enabling high-threshold predictions from moderate-threshold EEC measurements.

\end{enumerate}

\emph{Notation:} Boldface lowercase and uppercase letters denote vectors and matrices, respectively; $(\cdot)^T$ and $(\cdot)^H$ denote transpose and Hermitian transpose. We use $\mathbb E[\cdot]$ for expectation and $\mathcal{CN}(\boldsymbol\mu,\mathbf R)$ for a proper complex Gaussian distribution with mean $\boldsymbol\mu$ and covariance $\mathbf R$. The symbols $\sim$ and $\dot\sim$ mean ``distributed as'' and ``approximately distributed as,'' respectively.

\section{Signal Model}
\label{sec:signal_model}

\subsection{System Setup with Range-Gated Processing}

We consider a coherent distributed radar system consisting of $M$ spatially separated receiving stations and only one active transmitter during a coherent processing interval (CPI). This one-transmit–multi-receive setup can be implemented via time-division multiplexing such that, at any given CPI, only one station transmits while all stations listen. We do not consider multi-antenna arrays at each station, i.e., each station is equipped with a single transmit/receive channel. This simplification isolates the core difficulty of this work, which is distributed coherent integration under severe quantization and clutter, without introducing additional spatial-array degrees-of-freedom that are largely orthogonal to the proposed framework. The CPI contains $N$ pulses with pulse repetition interval (PRI) $T_r$. Let $s(t)$ denote the known complex baseband transmit pulse waveform, which lasts for a time interval of $t_0 \ (\leq T_r)$. Radar processing is assumed to be range-gated, namely, we focus on a fixed range gate. This is standard in practical detection systems, where coarse range can be obtained from tracking, prior scans, or parallel range processing. The main goal of this paper is to study the Doppler/velocity-domain detection at a fixed range gate, while range scanning can be performed by repeating the same procedure over multiple gates.

\subsection{Fast-Time Model and One-Bit Sampling}
We first introduce the scatterer-level time-domain echo model. 
Let $y_m^{p}(t)$ denote the complex baseband received signal at station $m\in\{1,\dots,M\}$ during the $p$-th pulse ($p=0,\dots,N-1$), observed over a fast-time window corresponding to the range gate. Using a linear superposition of delayed and Doppler-shifted echoes~\cite{richards2022radar}, we write
\begin{equation}
y_m^{p}(t) = e_m^{p}(t) + c_m^{p}(t) + w_m^{p}(t),
\end{equation} 
where $w_m^{p}(t)$ is the baseband noise, $e_m^{p}(t)$ is the target echo at station $m$, which is modeled as 
\begin{equation}
e_m^{p}(t)= \beta_{m,0}\, s(t-\tau_{m,0})\, e^{\imath  p\Omega  f_{m,0} },
\end{equation}
with $\Omega=2\pi T_r$, $\beta_{m,0}\in\mathbb{C}$ is the unknown complex scattering coefficient, $\tau_{m,0}$ is the propagation delay, and $f_{m,0}$ is the Doppler frequency at station $m$. %
The clutter echo $c_m^{p}(t)$ is modeled as a superposition of returns from a large number of scatterers~\cite{ward1990scattering}. At station $m$, it is represented by a sum of $K_m$ weak scatterers:
\begin{align}
c_m^{p}(t)=\sum_{k=1}^{K_m} \beta_{m,k}\, s(t-\tau_{m,k})\, e^{\imath  p \Omega f_{m,k}},
\end{align}
where each scatterer $k$ has its own complex coefficient $\beta_{m,k}$, delay $\tau_{m,k}$, and Doppler $f_{m,k}$.
Importantly, $\tau_{m,k}$ varies across scatterers, so many clutter returns only partially overlap with the fast-time window corresponding to the range gate. This  means that the effective fast-time waveform contributed by a scatterer is a truncated/weighted version of the delayed pulse within the observation window. Concretely, let $d_m(t)$ denote the indicator function of the range-gating window for station $m$: 
\begin{equation}
 d_m(t)= \begin{cases}
1, & \tau_{m,0}<t<\tau_{m,0}+t_0,\\
0, &\text{otherwise}. \\
\end{cases}
\end{equation}
Then, the observed fast-time signal is effectively $d_m(t)\,y_m^{p}(t)$, and each term becomes $d_m(t)s(t-\tau_{m,k})$, which remains linear in the scatterer coefficients $\beta_{m,k}$.
Let the fast-time sampling period be $T_s$, and let $\ell=0,\dots,L-1$ index samples within the gated window. Define the  gated, delayed waveform at station $m$ 
for delay $\tau$ as
\begin{equation}
\mathbf{s}_m(\tau)\triangleq \big[d_m(\ell T_s)\, s(\ell T_s-\tau)\big]_{\ell=0}^{L-1}\in\mathbb{C}^{L}.\label{eq:smtau}
\end{equation} 
Then the fast-time sample vector at station $m$ and pulse $p$  can be written compactly as
\begin{multline}
\mathbf{y}_m^p = \beta_{m,0}\, e^{\imath  p\Omega  f_{m,0}}\mathbf{s}_m(\tau_{m,0})\\
+\sum_{k=1}^{K_m}\beta_{m,k}\, e^{\imath p\Omega f_{m,k} }\mathbf{s}_m(\tau_{m,k}) + \mathbf{w}_m^p,\label{eq:fasttime_vec}
\end{multline}
where $\mathbf{w}_m^p\in\mathbb{C}^{L}$ is the sampled noise vector at station $m$.
In the regime considered here, $K_m$ is large and each $|\beta_{m,k}|$ is small.
To describe the fluctuations of these scattering coefficients, we adopt the compound-Gaussian representation~\cite{ward1990scattering,ollila2012compound} 
\begin{equation}
\beta_{m,k} = \sqrt{\xi_{m,k}}\, g_{m,k}, 
\label{compound_model}
\end{equation}
where $g_{m,k}\sim\mathcal{CN}(0,\sigma_g^2)$ and $\xi_{m,k}>0$ model the mutually independent speckle and texture components, respectively. The coefficients, delays, and Dopplers are fixed within a CPI, and pulse-to-pulse variation follows the waveform and Doppler phase. Conditioning on texture and scene parameters while averaging over Gaussian speckle gives a conditionally Gaussian clutter term. Independent Monte Carlo CPIs redraw these quantities in Section~\ref{sec:vanishing_quantization_loss}. Gamma texture gives a $K$-distributed scatterer amplitude~\cite{watts1990maritime}; inverse Gaussian texture provides another compound-Gaussian model~\cite{ollila2012compound}. Different texture laws change the mixing distribution~\cite{ing2016texture}. The vanishing-loss result in Section~\ref{sec:detector} does not depend on the specific texture distribution, provided that the weak-input and clutter-dominance assumptions hold. %

The one-bit quantizer is
\begin{align}
\mathcal{Q}(x)\triangleq \operatorname{sign}(\Re\{x\}) + \imath \operatorname{sign}(\Im\{x\}),\label{quantizer}
\end{align}
applied entrywise to the fast-time samples. The one-bit data at station $m$ and pulse $p$ are
\begin{align}
\mathbf{z}_m^p = \mathcal{Q}\big(\mathbf{y}_m^p\big)\in\{\pm 1 \pm \imath\}^{L}.
\end{align}
These one-bit samples $\mathbf{z}_m^p$ are then forwarded to the fusion center, where range/Doppler processing and distributed coherent detection are performed. %

\subsection{Matched Filtering and Slow-Time Model at Fixed Range Bin}

We apply a linear matched filter to the one-bit samples to coherently accumulate weak echoes. The Gaussian quantization moments underlying this approximation are described in~\cite{demir2021bussgang,li2017channel}. 
For a fixed reference delay $\tau_0$ (the range bin of interest), define the matched-filter template
$\mathbf{s}_m(\tau_0)$ as in \eqref{eq:smtau}. Using the one-bit samples, the range-compressed output
at station $m$ and pulse $p$ is
\begin{equation}
x_m[p] \triangleq \h_m^{H}\mathbf{z}_m^{p}, \qquad p=0,\ldots,N-1,
\label{eq:xbm_def}
\end{equation}
where $\h_m$ denotes the fast-time matched-filter weights, typically $\mathbf{s}_m(\tau_0)$, with any fast-time prewhitening absorbed into the weights.
Stacking over pulses yields the slow-time vector
\begin{equation}
\mathbf{x}_m=[x_m[0],\ldots,x_m[N-1]]^{T}\in\mathbb{C}^{N}.
\label{eq:slowtime_vector}
\end{equation}

Since $Q(\cdot)$ is applied entrywise and is nonlinear, substituting the scatterer-level model \eqref{eq:fasttime_vec} into the one-bit pulse compression \eqref{eq:xbm_def} does not yield an exact additive decomposition over the scatterers. 
Nevertheless, the statistics of the pulse-compressed output remain tractable because the fast-time observation is Gaussian after conditioning on the clutter realization.

To make this precise, we highlight the two layers of randomness in $\{x_m[p]\}$.
First, although the clutter realization is fixed within a CPI, the receiver noise $\mathbf{w}_m^p$ induces randomness in the one-bit samples and hence in $x_m[p]$.
Second, the clutter realization itself varies across CPIs due to random scatterer coefficients under the compound model \eqref{compound_model}%
. 
In this subsection, we isolate the effect of one-bit quantization by characterizing the conditional distribution of $x_m[p]$ given a fixed clutter realization. Specifically, let 
\begin{equation}
\mathcal{B}\triangleq\{(\beta_{m,k},\tau_{m,k},f_{m,k}):m=1,\ldots,M,\ k=1,\ldots,K_m\},
\end{equation}
collect the clutter scatterer parameters within the CPI. Conditioned on $\mathcal{B}$, the noiseless echo term in \eqref{eq:fasttime_vec} becomes deterministic, and thus $\mathbf{y}_m^p$ is a proper complex Gaussian vector since only the additive noise $\mathbf{w}_m^p$ remains random. 
We show next that, in the thermal-noise-dominated (low-SNR-per-sample) regime, one-bit quantization becomes {\it statistically linear}, which leads to a tractable effective slow-time model. The following lemma combines the weak-mean expansion with the Gaussian-input quantization moments~\cite{demir2021bussgang,li2017channel}. 

\begin{lemma}[One-bit matched-filter output: Gaussian characterization and low-SNR linearization] \label{lemma_linearity}
Let $\mathbf y\in\mathbb C^{L}$ be a proper complex Gaussian vector
\begin{equation}
\mathbf y \sim \mathcal{CN}(\boldsymbol{\mu}_y,\mathbf R_y),
\end{equation}
where $\mathbf R_y$ has equal marginal variances across fast-time samples, i.e.,
$\mathrm{diag}(\mathbf R_y)=\sigma^2\mathbf 1_L,$ while $\mathbf R_y$ itself is not restricted to be diagonal. Assume the weighted quantized samples satisfy a central-limit approximation, with no dominant sample and sufficiently weak dependence. Suppose $L$ is large and the per-sample ``instantaneous SNR'' is small such that
\begin{equation}
\varepsilon \triangleq \|\boldsymbol\mu_y\|_\infty/\sigma \ll 1 .
\end{equation}
Defining $\mathbf z=\mathcal Q(\mathbf y)$, and for a deterministic matched-filter template $\mathbf h\in\mathbb C^{L}$, the one-bit matched-filter output is
\begin{equation}
x \triangleq \mathbf h^{H}\mathbf z \, \dot\sim\ \mathcal{CN}\!\Big(\boldsymbol\mu_x,\ \mathbf h^{H}\mathbf R_z\mathbf h\Big),
\end{equation}
where the mean of the one-bit output admits the first-order approximation
\begin{align}
\boldsymbol\mu_x
&=\frac{2}{\sqrt{\pi}\sigma}\,\mathbf h^{H}\boldsymbol\mu_y + \mathcal O(\varepsilon^3).
\end{align}
To leading order in the weak mean, the covariance of $\mathbf z$ is given by the complex arcsine law:
\begin{align}\label{arcsin_law}
\mathbf R_z 
&\approx \mathcal{A}(\mathbf C)\Nn\\%
&=\frac{4}{\pi}\Big(\arcsin(\Re\{\mathbf C\})+\imath\,\arcsin(\Im\{\mathbf C\})\Big), 
\end{align}
where
$\mathbf C\triangleq \mathbf R_y/\sigma^2$ 
and $\arcsin(\cdot)$ is applied elementwise. Moreover, it is exact at $\boldsymbol\mu_y=\mathbf0$.
\end{lemma}

To apply Lemma~\ref{lemma_linearity}, let $\mathbf R_{y,m}\triangleq\operatorname{Cov}(\mathbf w_m^p)$ denote the receiver-noise covariance at station $m$. The receiver noise is independent of the clutter realization $\mathcal B$. Conditioning on $\mathcal B$ makes the target and clutter echoes deterministic, so that $\operatorname{Cov}(\mathbf y_m^p\mid\mathcal B)=\mathbf R_{y,m}$. 
In practical radar receivers, this fast-time baseband noise is wide-sense stationary over the observation window after fixed analog front-end filtering and uniform sampling. 
Therefore, although the samples may be temporally correlated due to, e.g., bandlimiting or oversampling, their marginal variances are identical, so that the covariance matrix $\mathbf R_{y,m}$ satisfies
$\mathrm{diag}(\mathbf R_{y,m})=\sigma_m^2\mathbf 1_L$. 

When the weak-input and weighted-sum conditions of Lemma~\ref{lemma_linearity} also hold, using $\mathbf{y}=\mathbf{y}_m^p$ and $\mathbf{h}=\mathbf{h}_m$ yields the approximate distribution
\begin{equation}
x_m[p]|\mathcal{B} \ \dot\sim\ \mathcal{CN}\!\big(\mu_{x,m}[p],\ \mathbf h_m^{H}\mathbf R_{z,m}\mathbf h_m\big),
\label{eq:xmp_gauss}
\end{equation}
where $\mathbf R_{z,m}\approx\mathcal{A}(\mathbf C_m)$,  with 
$\mathbf C_m = \mathbf R_{y,m}/\sigma_m^2$ being the noise correlation matrix at station $m$, and the mean admits the weak signal approximation
\begin{equation}
\mu_{x,m}[p]
\approx \frac{2}{\sqrt{\pi}\sigma_m}\ \mathbf h_m^{H}\boldsymbol{\mu}_{y,m}^{p}.
\label{eq:muxmp}
\end{equation}
Under $\mathcal H_0$, the mean component is the only one that contains clutter echo within the gated fast-time window, so that
\begin{equation}
\mu_{x,m}[p]
=
\sum_{k=1}^{K_m}\frac{2}{\sqrt{\pi}\sigma_m}\beta_{m,k}\, e^{\imath p\Omega f_{m,k} }\mathbf{h}_m^H\mathbf{s}_m(\tau_{m,k}).
\label{eq:u_H0}
\end{equation}
Stacking over $p$ yields the slow-time vector form:
\begin{align}
\bmu_{x,m}^0=\sum_{k=1}^{K_m}\alpha_{m,k}\mathbf{a}(f_{m,k}),
\end{align}
where the unit-norm Doppler steering vector and its associated amplitude are defined by
\begin{align}
\mathbf{a}(f)=&\frac{1}{\sqrt N}[1,e^{\imath \Omega f},\ldots,e^{\imath \Omega f(N-1)}]^{T},\label{eq:doppler_steering}\\
\alpha_{m,k}=& \frac{2\sqrt N}{\sqrt{\pi}\sigma_m} \beta_{m,k}\, \mathbf{h}_m^H\mathbf{s}_m(\tau_{m,k}).
\end{align}
Similarly, under $\mathcal H_1$, we have
\begin{equation}
\bmu_{x,m}^1=\alpha_{m,0}\mathbf{a}(f_{m,0})+\bmu_{x,m}^0.
\end{equation}
Assuming independent receiver noise across pulses, the conditional covariance of the stacked output is, to leading order,
\begin{equation}
\operatorname{Cov}(\mathbf x_m\mid\mathcal B)\approx\mathbf R_{x,m}\triangleq(\mathbf h_m^H\mathbf R_{z,m}\mathbf h_m)\mathbf I_N.
\label{eq:conditional_slowtime_covariance}
\end{equation}
\noindent Finally, the detection problem at a fixed range bin can be written as
\begin{equation}
\label{eq:detection_problem}
\begin{array}{l}
\mathcal H_0:\ \mathbf x_m|\mathcal{B} \ \dot\sim\ \mathcal{CN}(\bmu_{x,m}^0,\mathbf R_{x,m}),\\
\mathcal H_1:\ \mathbf x_m|\mathcal{B} \ \dot\sim\ \mathcal{CN}(\bmu_{x,m}^1,\mathbf R_{x,m}). 
\end{array}
\end{equation}

\section{Vanishing Loss}
\label{sec:detector}

Our objective in this section is to identify a clutter-dominated regime in which one-bit quantization loss can vanish and to explain the mechanism behind this behavior. Such vanishing loss contrasts with the classical $2$~dB penalty reported for one-bit detection in purely thermal-noise-limited models. To isolate the effect of quantization, we compare the one-bit and high-resolution sampling models for fixed target signatures and known interference covariances within the effective Gaussian model.

\subsection{$\infty$-Bit Baseline}
As a baseline, consider an ideal high-resolution ADC and define the unquantized pulse-compressed output at station $m$ and pulse $p$ as
\begin{equation}
\tilde{x}_m[p] \triangleq \mathbf{h}_m^H \mathbf{y}_m^p .
\end{equation}
Conditioned on the clutter realization $\mathcal{B}$, $\mathbf{y}_m^p$ is proper complex Gaussian with mean $\boldsymbol\mu_{y,m}^p$ and covariance $\mathbf{R}_{y,m}=\sigma_m^2\mathbf{C}_m$ (with $\mathrm{diag}(\mathbf{C}_m)=\mathbf{1}$). Hence,
\begin{equation}
\tilde{x}_m[p]\mid\mathcal{B} \sim \mathcal{CN}\!\big(\tilde{\mu}_{x,m}[p],\, \tilde{\sigma}_{x,m}^2\big),
\end{equation}
with
\begin{align}
\tilde{\mu}_{x,m}[p] &= \mathbf{h}_m^H\boldsymbol\mu_{y,m}^p, &
\tilde{\sigma}_{x,m}^2 &= \mathbf{h}_m^H\mathbf{R}_{y,m}\mathbf{h}_m.
\label{eq:highres_moments}
\end{align}
Stacking the outputs as in~\eqref{eq:slowtime_vector} gives $\tilde{\mathbf{x}}_m$.
Using the scatterer-level model in \eqref{eq:fasttime_vec} and defining 
\begin{align}
\tilde{\alpha}_{m,k}\triangleq \sqrt N\,\beta_{m,k}\mathbf{h}_m^H\mathbf{s}_m(\tau_{m,k}),
\label{eq:highres_alpha}
\end{align}
we obtain the slow-time representation
\begin{equation}
\tilde{\mathbf{x}}_m = \tilde{\mathbf{s}}_m + \tilde{\mathbf{c}}_m + \tilde{\mathbf{n}}_m,
\label{eq:highres_slowtime_model}
\end{equation}
where 
\begin{align}
\tilde{\mathbf{c}}_m  &=\sum_{k=1}^{K_m}\tilde{\alpha}_{m,k}\mathbf{a}(f_{m,k}), &
\tilde{\mathbf{s}}_m  &=\tilde{\alpha}_{m,0}\mathbf{a}(f_{m,0})
\label{eq:highres_components}
\end{align}
with $\tilde{\alpha}_{m,0}=0$ under $\mathcal{H}_0$.
The receiver noise is assumed independent of the clutter
parameters $\mathcal B$ and independent across pulses.
For the fixed matched filter $\mathbf h_m$, the resulting
slow-time noise therefore satisfies
\begin{equation}
\tilde{\mathbf n}_m
\sim \mathcal{CN}
\bigl(\mathbf 0,\tilde{\sigma}_{x,m}^2\mathbf I_N\bigr).
\label{eq:highres_noise}
\end{equation}
\subsection{Vanishing Loss in Clutter-Limited Regime}

We now formalize the \emph{vanishing loss} statement. The key point is that the two conditions used below refer to {different processing stages}. 
First, the low-SNR condition in Lemma~\ref{lemma_linearity} is imposed on the {raw fast-time samples} at the ADC input: each target/clutter echo sample is weak relative to the receiver noise, so the one-bit quantizer operates in its statistical linear region. 
Second, after pulse compression, the weak echoes are coherently accumulated over the fast-time window and over many clutter scatterers, so that the resulting slow-time clutter term may become much stronger than the post-compression noise term. 

Using the high-resolution decomposition~\eqref{eq:highres_slowtime_model} and Lemma~\ref{lemma_linearity}, in the low-SNR-per-sample regime, the one-bit slow-time vector can be represented, up to higher-order terms, as a scaled version of the ideal output plus an additional distortion term, consistent with the Gaussian-input Bussgang decomposition~\cite{demir2021bussgang}:
\begin{equation}
\mathbf{x}_m = \kappa_m \tilde{\mathbf{x}}_m + \mathbf{r}_m,
\end{equation}
where $\kappa_m=\tfrac{2}{\sqrt{\pi}\sigma_m}$ for the %
complex one-bit quantizer, and $\mathbf{r}_m$ is the residual quantization-distortion term whose mean is negligible and whose covariance is induced by the difference between the arcsine-law covariance and the linearly scaled covariance.

It is convenient to rescale the one-bit slow-time vector by the known factor $\kappa_m$:
\begin{align}
\bar{\mathbf{x}}_m &\triangleq \kappa_m^{-1}\mathbf{x}_m 
= \tilde{\mathbf{s}}_m+\tilde{\mathbf{c}}_m+\bar{\mathbf{n}}_m,
&
\bar{\mathbf{n}}_m &\triangleq \tilde{\mathbf{n}}_m+\kappa_m^{-1}\mathbf{r}_m.
\label{eq:rescaled_onebit_model}
\end{align}
Thus, after rescaling, the one-bit and high-resolution models share the {same} target term $\tilde{\mathbf{s}}_m$ and the {same} clutter term $\tilde{\mathbf{c}}_m$, and they differ only in the effective additive-noise term.

From~\eqref{eq:highres_moments} and~\eqref{eq:highres_noise}, the unquantized matched-filtered noise has conditional covariance $\mathbf{R}_{\tilde{n},m}=\tilde{\sigma}_{x,m}^{2}\mathbf{I}_N$, whereas the rescaled one-bit effective noise has covariance
\begin{equation}
\mathbf{R}_{\bar{n},m}
= \kappa_m^{-2}\mathbf{R}_{x,m}
= \frac{\pi\sigma_m^2}{4}\big(\mathbf{h}_m^{H}\mathbf{R}_{z,m}\mathbf{h}_m\big)\mathbf{I}_N,
\label{eq:Rbar_n}
\end{equation}
with $\mathbf{R}_{z,m}$ given by the complex arcsine law in Lemma~\ref{lemma_linearity}.
For white fast-time noise, \eqref{eq:Rbar_n} reduces to the classical noise-inflation factor $\pi/2$, corresponding to the well-known $2$~dB loss in noise-limited one-bit detection.

The clutter term, however, behaves differently. For the clutter in~\eqref{eq:highres_components}, with coefficients defined in~\eqref{eq:highres_alpha}, conditioning on texture, delays, and Dopplers while averaging over speckle gives the covariance
\begin{equation}
\mathbf{R}_{c,m}\triangleq \mathbb{E}\!\left[\tilde{\mathbf{c}}_m\tilde{\mathbf{c}}_m^{H}\right].
\label{eq:Rc_def}
\end{equation}
If the scatterer coefficients are uncorrelated, $\mathbf{R}_{c,m}$ becomes
\begin{equation}
\mathbf{R}_{c,m}
= N\sum_{k=1}^{K_m}\mathbb{E}\!\left[|\beta_{m,k}|^2\right]
\big|\mathbf{h}_m^H\mathbf{s}_m(\tau_{m,k})\big|^2
\,\mathbf{a}(f_{m,k})\mathbf{a}^{H}(f_{m,k}).
\label{eq:Rc_expand}
\end{equation}
Equation \eqref{eq:Rc_expand} shows that the clutter covariance increases with both the number of scatterers and the pulse-compression gains 
$\big|\mathbf{h}_m^H\mathbf{s}_m(\tau_{m,k})\big|^2$, whereas the additive-noise covariance in \eqref{eq:Rbar_n} does not grow with $K_m$.
Therefore, even though the raw fast-time samples are noise-dominated at the ADC input, the pulse-compressed output can be clutter-dominated.

To quantify this regime, define the post-compression clutter-to-noise ratios
\begin{align}
\Gamma_m^{(\infty)} \triangleq \frac{\mathrm{tr}(\mathbf{R}_{c,m})}{\mathrm{tr}(\mathbf{R}_{\tilde{n},m})},
\qquad
\Gamma_m^{(1)} \triangleq \frac{\mathrm{tr}(\mathbf{R}_{c,m})}{\mathrm{tr}(\mathbf{R}_{\bar{n},m})}.
\end{align}
The noise covariances $\mathbf R_{\tilde n,m}$ and $\mathbf R_{\bar n,m}$ differ only by a constant factor. Thus, if either $\Gamma_m^{(\infty)}$ or $\Gamma_m^{(1)}$ tends to infinity, so does the other. However, these ratios compare total clutter power with total noise power. For example, strong clutter concentrated around a few Doppler frequencies can make these ratios large, while noise remains significant at other frequencies. The following proposition therefore uses a stronger covariance condition: the additive noise must become negligible relative to clutter for all linear combinations of the slow-time samples.

\begin{proposition}[Vanishing one-bit loss in clutter]
Assume: (i) the low-SNR-per-sample condition in Lemma~\ref{lemma_linearity} holds, so that the one-bit pulse-compressed model \eqref{eq:rescaled_onebit_model} is accurate; 
(ii) $\mathbf R_{c,m}\succ\mathbf 0$ and both additive-noise covariances satisfy
\begin{equation}
\epsilon_m\triangleq
\max_{b\in\{\infty,1\}}
\left\|\mathbf R_{c,m}^{-1/2}\mathbf R_{n,m}^{(b)}\mathbf R_{c,m}^{-1/2}\right\|_2\to0,
\end{equation}
where $\mathbf R_{n,m}^{(\infty)}=\mathbf R_{\tilde n,m}$ and $\mathbf R_{n,m}^{(1)}=\mathbf R_{\bar n,m}$.  
Within this Gaussian-equivalent model, the fixed-signature detection performances with known covariances converge at a fixed false-alarm probability. In particular, the required SCR gap approaches $0$~dB.
\end{proposition}

\begin{IEEEproof}
For a fixed target signature and known covariance, detectability under the Gaussian model is characterized by the signal energy after whitening~\cite{kay1998detection,scharf1994matched}. Using $\epsilon_m$ as defined above, set
\begin{align}
\delta_m^{(b)}&=\tilde{\mathbf s}_m^H(\mathbf R_{c,m}+\mathbf R_{n,m}^{(b)})^{-1}\tilde{\mathbf s}_m,\\
\delta_{c,m}&=\tilde{\mathbf s}_m^H\mathbf R_{c,m}^{-1}\tilde{\mathbf s}_m.
\end{align}
Whitening by $\mathbf R_{c,m}$ bounds the eigenvalues of the normalized inverse between $(1+\epsilon_m)^{-1}$ and $1$, so
\begin{equation}
\frac{\delta_{c,m}}{1+\epsilon_m}\le\delta_m^{(b)}\le\delta_{c,m}.
\end{equation}
For fixed $M$ and $N$, the two summed noncentralities therefore have ratio tending to one. Their fixed-false-alarm receiver operating characteristic (ROC) curves, and the required target-power ratios at a fixed detection probability, coincide asymptotically.
\end{IEEEproof}

\begin{remark}[Why the classical 2 dB loss disappears]
In thermal-noise-limited detection, the additive noise is the dominant interference term, so the one-bit noise inflation in \eqref{eq:Rbar_n} directly yields the familiar $2$~dB loss. 
In clutter-limited detection, by contrast, the dominant randomness is the clutter term $\tilde{\mathbf{c}}_m$, which undergoes the same low-SNR scaling as the target term. 
Hence the target-to-clutter ratio is preserved, and the one-bit performance loss vanishes.
\end{remark}

\section{Velocity-Plane Detector for One-Bit Data}
\label{sec:velocity_search}

Having identified the conditions for vanishing quantization loss, we now turn our attention to detection with unknown target velocity and station amplitudes. We use the same effective slow-time model to reformulate the joint velocity-search detector of~\cite{he2010homogeneous}.

For the one-bit matched-filter outputs defined in~\eqref{eq:xbm_def}, the weak-input and weighted-sum conditions of Lemma~\ref{lemma_linearity} give
\begin{equation}
\mathbb E[x_m[p]\mid\mathcal B]
\approx\kappa_m\mathbf h_m^H\boldsymbol\mu_{y,m}^p,
\label{eq:onebit_scan_linearization}
\end{equation}
where $\kappa_m=2/(\sqrt{\pi}\sigma_m)$. With pulse-to-pulse independent receiver noise, stacking the outputs gives the representation as:
\begin{equation}
\mathbf x_m \approx\kappa_m\big(\tilde{\mathbf s}_m+\tilde{\mathbf c}_m+\bar{\mathbf n}_m\big),\\
\label{eq:onebit_scan_decomposition}
\end{equation}
where $\bar{\mathbf n}_m\mid\mathcal B \ \dot\sim\ \mathcal{CN}(\mathbf0,\mathbf R_{\bar n,m})$, the effective-noise covariance $\mathbf R_{\bar n,m}$ is given by~\eqref{eq:Rbar_n} and is independent of the clutter realization $\mathcal B$ to leading order.

Conditioning on textures, delays, and Dopplers while marginalizing over Gaussian speckle gives the clutter distribution and total interference covariance
\begin{align}
\tilde{\mathbf c}_m&\sim\mathcal{CN}(\mathbf0,\mathbf R_{c,m}), &
\mathbf R_m&\approx\kappa_m^2\big(\mathbf R_{c,m}+\mathbf R_{\bar n,m}\big),
\label{eq:onebit_scan_covariance}
\end{align}
where $\mathbf R_{c,m}$ is defined in~\eqref{eq:Rc_def}. Thus, under the effective Gaussian approximation, the detection problem becomes
\begin{equation}
\begin{aligned}
\mathcal H_0:&\quad \mathbf x_m\ \dot\sim\ \mathcal{CN}(\mathbf 0,\mathbf R_m),\\
\mathcal H_1:&\quad \mathbf x_m\ \dot\sim\ \mathcal{CN}\!\left(\alpha_{m,0}\mathbf a_m(\mathbf v),\mathbf R_m\right).
\end{aligned}
\label{eq:onebit_scan_model}
\end{equation}
Here, $\alpha_{m,0}=\kappa_m\tilde\alpha_{m,0}$ is the target amplitude under the unit-norm convention in~\eqref{eq:doppler_steering}, including the quantization attenuation and pulse-compression response. The generalized likelihood ratio test (GLRT) below assumes conditionally independent station disturbances and known positive-definite $\mathbf R_m$, fixed across hypotheses and candidate velocities.

For a candidate planar velocity $\mathbf v\in\mathbb R^2$, the known transmitter--receiver geometry determines
\begin{equation}
f_m(\mathbf v)=\frac{2}{\lambda}\mathbf q_m^T\mathbf v,
\label{eq:doppler_mapping}
\end{equation}
where $\lambda$ is the carrier wavelength and $\mathbf q_m$ is half the sum of the transmit and receive line-of-sight unit vectors. Using~\eqref{eq:doppler_steering}, the associated slow-time steering vector is
\begin{equation}
\mathbf a_m(\mathbf v)=\mathbf a\big(f_m(\mathbf v)\big).
\label{eq:steering_vector}
\end{equation}

Let $\boldsymbol\alpha=[\alpha_{1,0},\ldots,\alpha_{M,0}]^T$ collect the unknown station amplitudes. For fixed $\mathbf v$, the model log-likelihood ratio is
\begin{multline}
\ell(\boldsymbol\alpha,\mathbf v)=\sum_{m=1}^{M}\Big[
2\Re\!\left\{\alpha_{m,0}^*\mathbf a_m^H\mathbf R_m^{-1}\mathbf x_m\right\}\\
-|\alpha_{m,0}|^2\mathbf a_m^H\mathbf R_m^{-1}\mathbf a_m\Big],
\label{eq:onebit_log_likelihood}
\end{multline}
where $\mathbf a_m=\mathbf a_m(\mathbf v)$. Maximization over the complex amplitudes gives $\widehat\alpha_{m,0}=(\mathbf a_m^H\mathbf R_m^{-1}\mathbf x_m)/(\mathbf a_m^H\mathbf R_m^{-1}\mathbf a_m)$ and hence~\cite{kay1998detection,scharf1994matched}
\begin{align}
\eta_m(\mathbf v)&\triangleq
\frac{\mathbf a_m(\mathbf v)^H\mathbf R_m^{-1}\mathbf x_m}
{\sqrt{\mathbf a_m(\mathbf v)^H\mathbf R_m^{-1}\mathbf a_m(\mathbf v)}},
\label{eq:gmf_output}\\
T(\mathbf v)&\triangleq\max_{\boldsymbol\alpha}\ell(\boldsymbol\alpha,\mathbf v)
=\sum_{m=1}^{M}|\eta_m(\mathbf v)|^2.
\label{eq:scan_statistic}
\end{align}
Here, $\eta_m(\mathbf v)$ denotes the output of the generalized matched filter (GMF).
Thus, the detector sums station-wise coherent matched-filter energies. For a continuous search region $\mathcal V$ and a dense grid $\mathcal V_{\rm grid}\subset\mathcal V$, respectively, it uses
\begin{align}
T_{\max}^{\rm c}&\triangleq\sup_{\mathbf v\in\mathcal V}T(\mathbf v),
\label{eq:continuous_search}\\
T_{\max}&\triangleq\max_{\mathbf v\in\mathcal V_{\rm grid}}T(\mathbf v)
\mathop{\gtrless}_{\mathcal H_0}^{\mathcal H_1}\tau.
\label{eq:grid_search}
\end{align}
In practice, the detector uses the grid statistic $T_{\max}$. Its continuous counterpart $T_{\max}^{\rm c}$ is introduced for the EEC-based false-alarm analysis and threshold calibration in Section~\ref{sec:null distribution}. A sufficiently dense grid approximates this continuous search.

\section{EEC-Based CFAR Calibration}
\label{sec:null distribution}

Under $\mathcal H_0$, the Gaussian-equivalent model established above allows the velocity-plane scan statistic $T(\mathbf{v})$ to be treated as a smooth $\chi^2$-type random field over $\mathcal{V}$. Constant false alarm rate (CFAR) calibration requires the tail probability of its maximum over the velocity search region. We therefore employ the EEC framework~\cite{adler2007randomfields,taylor2005validity} to approximate this extreme-value probability through the excursion set
\begin{equation}%
A_\tau=\{\mathbf v\in\mathcal V:T(\mathbf v)\ge \tau\}.
\end{equation}
At high thresholds, the excursion set $A_\tau$ typically consists of a small number of isolated components around local maxima and rarely contains holes. 
In two dimensions, the EC of a set, denoted by $\chi(\cdot )$, is the number of connected components minus the number of holes.
Therefore, $\chi(A_\tau)$ is approximately the number of high-level excursion components. We define its expectation
\begin{equation}
\mu(\tau)\triangleq \mathbb E[\chi(A_\tau)]
\label{eq:eec_def}
\end{equation}
as the EEC of the excursion set, which approximately represents the expected number of threshold-exceedance clusters at high thresholds.

For a two-dimensional velocity search region, the EEC comprises contributions of three intrinsic dimensions: a zero-dimensional pointwise contribution term associated with single-point exceedance probability, a one-dimensional boundary contribution associated with excursion
components intersecting the boundary, and a two-dimensional
interior contribution associated with excursion components
formed in the interior.
Specifically, the EEC decomposition has three contribution terms with the form
\begin{equation}
L_d(\mathcal V)\rho_d(\tau),\qquad d=0,1,2.
\end{equation}

Here, $\rho_d(\tau)$ is the $d$-th EC density, representing the expected local contribution to the EC per unit $d$-dimensional intrinsic measure at threshold $\tau$, whereas $L_d(\mathcal V)$ is the corresponding Lipschitz–Killing curvature (LKC)~\cite[Chs.~6 and~10]{adler2007randomfields}. Moreover, $L_0(\mathcal V)$, $L_1(\mathcal V)$, and $L_2(\mathcal V)$ represent the EC of the search region, one half of its induced boundary length, and its induced area, respectively. Thus, the EC densities describe the threshold-dependent local excursion behavior, while the LKCs characterize the effective search geometry.

The intrinsic geometry is determined by the local correlation structure of the scan field rather than by ordinary Euclidean velocity distance. Concretely, the field-induced metric characterizes how rapidly the scan field decorrelates under a local velocity displacement, including its directional dependence. In Cartesian velocity coordinates, its local metric is represented by a symmetric positive-definite matrix $\mathbf G(\mathbf v)$, and an infinitesimal velocity displacement $d\mathbf{v}$ is assigned the squared field-induced length
\begin{equation}\label{length}
d\ell_G^2 = d\mathbf{v}^{T}\mathbf{G}(\mathbf{v})d\mathbf{v}.
\end{equation}
The metric is normalized such that $\frac{1}{2}d\ell_G^2$ equals the leading second-order aggregate covariance loss caused by $d\mathbf{v}$. 
The corresponding induced area element is 
\begin{equation}\label{area}
dA_G =
\sqrt{\det \G(\mathbf{v})} dv_x dv_y.
\end{equation}
Here, $dv_x$ and $dv_y$ are the coordinate differentials in the velocity plane.
These two elements determine the boundary-length and area terms entering $L_1(\mathcal V)$ and $L_2(\mathcal V)$, respectively.

To summarize, the following lemma elaborates the
EEC--Poisson approximation and the Kac-Rice representations used to evaluate
these quantities~\cite[Ch.~6]{azais2009levelsets}. The classical Gaussian kinematic formula assumes identically distributed Gaussian component fields~\cite{taylor2006gkf}. However, different station Doppler directions do not generally meet that requirement. We therefore state the local-density conditions explicitly and evaluate the resulting approximation for the geometries used here.

\begin{lemma}[EEC approximation for a two-dimensional smooth $\chi^2$-type field]
\label{lem:EED_approx}
Let $T(\mathbf{v})$, $\mathbf{v}\in\mathcal{V}\subset\mathbb{R}^2$, be a
sufficiently smooth $\chi^2$-type random field over a rectangle, with constant positive-definite field-induced metric $\mathbf G$. Assume the regularity and nondegeneracy needed for the Kac--Rice representations below. Suppose the metric-normalized local densities are spatially homogeneous and the one-dimensional density is the same along the boundary directions of this rectangle.
With the additional approximation that rare excursion clusters have a Poisson count, the high-threshold tail probability is modeled by
\begin{equation}
P_{\mathrm{FA}}(\tau)
=
\mathbb P\{A_\tau\neq\emptyset\}
\approx
1-\exp[-\mu(\tau)],
\label{eq:eec_poisson}
\end{equation}
where $\mu(\tau)$ is defined in \eqref{eq:eec_def} and can be approximately decomposed as
\begin{equation}
\mu(\tau)
\approx
L_0(\mathcal V)\rho_0(\tau)
+
L_1(\mathcal V)\rho_1(\tau)
+
L_2(\mathcal V)\rho_2(\tau).
\label{eq:gkf_2d}
\end{equation}
The dimensions $0,1,2$ exhaust the geometric terms for a planar domain. The approximation concerns the local-density factorization and the field model, rather than omitted higher dimensions. The Poisson conversion in~\eqref{eq:eec_poisson} is a separate approximation, checked numerically in Section~\ref{sec:eec_simulation}.
The EC densities admit the Kac–Rice representations
\begin{align}
\rho_0(\tau) &= \mathbb{P}\{T(\mathbf{v}) \ge \tau\}, \\[2mm]
\rho_1(\tau) &= p_T(\tau)\,
\mathbb{E}\left[
\left(\frac{\partial T}{\partial \xi_1}\right)_+
\,\middle|\, T=\tau
\right], \\[2mm]
\rho_2(\tau) &= \int_{\tau}^{\infty}
p_{T,\nabla_{\xi}T}(u,\mathbf{0})\,
\Nn\\
&\times \mathbb{E}\left[
\det\left(-\nabla_{\xi}^{2}T\right)
\,\middle|\,
T=u,\nabla_{\xi}T=\mathbf{0}
\right]\,du.
\end{align}
The associated LKCs are
\begin{align}
L_0(\mathcal V)&=\chi(\mathcal V),
\label{eq:L0_def}\\
L_1(\mathcal V)
&=
\frac{1}{2}
\int_{\partial\mathcal V}
\sqrt{\boldsymbol{\gamma}'(s)^T\mathbf G(\boldsymbol{\gamma}(s))\boldsymbol{\gamma}'(s)}\,ds,
\label{eq:L1_def}\\
L_2(\mathcal V)
&=
\int_{\mathcal V}
\sqrt{\det(\mathbf G(\mathbf v))}\,d\mathbf v.
\label{eq:L2_def}
\end{align}
Here, $(x)_+ \triangleq \max(x,0)$. The coordinate $\xi_1$ is locally
parameterized at unit speed with respect to the induced metric along a
one-dimensional boundary stratum, while $\boldsymbol{\xi}=(\xi_1,\xi_2)$
denotes a local coordinate system orthonormal with respect to
$\mathbf{G}(\mathbf{v})$. The functions $p_T$ and $p_{T,\nabla_{\xi}T}$
denote the marginal density of $T$ and the joint density of $T$ and its
metric-normalized gradient, respectively. Moreover,
$\gamma:[0,P]\to \partial\mathcal{V}$ is a piecewise $C^1$ parametrization
of the boundary by Euclidean arc length, such that $\|\gamma'(s)\|_2 = 1$.
\end{lemma}

The quantities and model-specific conditions required by Lemma~\ref{lem:EED_approx} are addressed in the following subsections. Subsection V-A derives the zeroth-order EC density from the pointwise null distribution. Subsection V-B derives the field-induced metric and the associated LKCs. The resulting covariance representation also establishes the spatial homogeneity of the metric-normalized local statistics. Subsection V-C derives and calibrates the higher-order EC densities, while assessing the remaining tangent-direction condition. With these quantities and conditions specified, the false-alarm probability is evaluated as
\begin{align}
P_{\mathrm{FA}}(\tau)
\approx
1-\exp\left[
-\sum_{d=0}^{2} L_d(\mathcal{V})\rho_d(\tau)
\right].
\label{pfa_1}
\end{align}

\subsection{Pointwise Null Distribution and Zeroth-Order EC Density}

We first characterize the scan statistic defined in \eqref{eq:scan_statistic} under $\mathcal H_0$. 
For a fixed candidate velocity $\mathbf v$,~\eqref{eq:gmf_output} gives
\begin{equation}
\eta_m(\mathbf v)\sim \mathcal{CN}(0,1), \qquad m=1,\ldots,M,
\label{eq:eta_null}
\end{equation}
provided that $\mathbf x_m\sim\mathcal{CN}(\mathbf 0,\mathbf R_m)$ under $\mathcal H_0$. 
Moreover, the station-wise disturbances are modeled as conditionally independent, and hence 
$\{\eta_m(\mathbf v)\}_{m=1}^{M}$ are independent for each fixed $\mathbf v$. 
Therefore, for any fixed $\mathbf v$,
\begin{align}
T(\mathbf v) &= \sum_{m=1}^{M}|\eta_m(\mathbf v)|^2
\sim \mathrm{Gamma}(M,1).
\label{eq:pointwise_gamma}
\end{align}
Therefore, the single-point exceedance probability at a threshold $\tau$ is
\begin{equation}\label{eq:rho0_general}
\rho_0(\tau)
\triangleq
\mathbb P\{T(\mathbf v)\ge \tau\}
=
e^{-\tau}\sum_{k=0}^{M-1}\frac{\tau^k}{k!}.
\end{equation}
According to Lemma~\ref{lem:EED_approx},~\eqref{eq:rho0_general} is the zeroth-order EC density required in the EEC decomposition. It is determined entirely by the pointwise null distribution and is independent of the geometry and size of the velocity search region.
As $\mathbf v$ varies, the same slow-time data are projected onto Doppler templates whose frequencies vary smoothly with the velocity hypothesis. Under the affine velocity-to-Doppler mapping adopted in this paper, the normalized station fields $\eta_m(\mathbf v)$, and hence $T(\mathbf v)$, vary smoothly over $\mathcal V$. Thus, $T(\mathbf v)$ forms a smooth $\chi^2$-type random field, providing the random-field structure required by Lemma~\ref{lem:EED_approx}. Its spatial correlation structure, which determines the field-induced metric and the higher-dimensional EEC contributions, is characterized next.

\subsection{Induced Metric and LKCs for Velocity Plane}
\label{sec:induced_metrics}

As the GMF handles known colored slow-time interference by whitening both observations and Doppler templates, its correlation is determined by the normalized whitened templates. For explicit metric and LKC expressions, we consider $\mathbf R_m=\nu_m^2\mathbf I$; for a general known covariance, the correlation and metric are obtained using the corresponding normalized pre-whitened Doppler templates. We use the unit-norm templates in~\eqref{eq:doppler_steering} and~\eqref{eq:steering_vector}, with uniformly spaced slow-time samples $t_p=pT_r$, $p=0,\ldots,N-1$, and a rectangular window. Under $\mathcal H_0$, the GMF in~\eqref{eq:gmf_output} gives the correlation between nearby velocity hypotheses as
\begin{align}
\mathbb{E}\!\left[
\eta_m(\mathbf{v}) \eta_m^{*}(\mathbf{v}+\Delta\mathbf{v})
\right]
&=
\mathbf{a}^{\mathrm{H}}\bigl(f_m(\mathbf{v})\bigr)
\mathbf{a}\bigl(f_m(\mathbf{v}+\Delta\mathbf{v})\bigr)  \Nn\\
&= r(\Delta f_m),
\end{align}
where
\begin{equation}
\Delta f_m \triangleq f_m(\mathbf{v}+\Delta\mathbf{v}) - f_m(\mathbf{v}) .
\end{equation}
and
\begin{equation}
r(\Delta f)
=
\frac{1}{N}
\sum_{p=0}^{N-1} e^{\imath 2\pi \Delta f t_p}.
\label{eq:freq_corr}
\end{equation}
Hence, the local correlation of the scan field is determined by the mismatch between the Doppler templates generated by neighboring velocity hypotheses.

The absolute phase associated with the choice of the slow-time origin does not affect the station statistic $|\eta_m(\mathbf{v})|^2$. We therefore define the centered slow-time samples $\tilde{t}_p=t_p-\bar{t}$, where $\bar{t}=(N-1)T_r/2$ is the mean sampling time. The corresponding phase-centered Doppler template and station field are
\begin{equation}
\begin{aligned}
\tilde{\mathbf{a}}(f)&=e^{-\imath 2\pi f\bar t}\mathbf{a}(f),\\
\tilde{\eta}_m(\mathbf{v})&=e^{\imath 2\pi f_m(\mathbf{v})\bar{t}}\eta_m(\mathbf{v}).
\end{aligned}
\end{equation}
Since $|\tilde{\eta}_m(\mathbf{v})|^2=|\eta_m(\mathbf{v})|^2$, phase centering leaves $T(\mathbf{v})$ unchanged while removing the component of the local template variation corresponding only to a deterministic phase rotation. The centered Doppler-domain covariance is
\begin{equation}
r_c(\Delta f)
= \tilde{\mathbf{a}}^{\mathrm{H}}(f)
   \tilde{\mathbf{a}}(f+\Delta f) 
= \frac{1}{N}
\sum_{p=0}^{N-1} e^{\imath 2\pi \Delta f (t_p-\bar{t})}.
\end{equation}
Since $\sum_{p=0}^{N-1}(t_p-\bar{t})=0$, the first-order term of $r_c(\Delta f)$ vanishes at $\Delta f=0$. Its second-order expansion is therefore
\begin{equation}\label{eq:corr_expansion}
r_c(\Delta f)
=
1-\frac{1}{2}\lambda_{\mathrm{eff}}(\Delta f)^2
+o((\Delta f)^2),
\end{equation}
where
\begin{equation}
\lambda_{\mathrm{eff}}
=
\frac{(2\pi)^2}{N}\sum_{p=0}^{N-1}(t_p-\bar t)^2
=\frac{(2\pi T_r)^2(N^2-1)}{12}.
\label{eq:lambda_eff}
\end{equation}
Thus, $\lambda_{\mathrm{eff}}$ characterizes the local sensitivity of the slow-time Doppler template to a small Doppler mismatch.

To map this Doppler-domain correlation structure to the velocity plane, define
\begin{equation}
\mathbf{d}_m \triangleq \nabla_{\mathbf{v}} f_m(\mathbf{v}),
\end{equation}
as the velocity-to-Doppler sensitivity vector of station $m$. Because the Doppler model adopted in this paper is affine in $\mathbf v$, $\mathbf d_m$ is independent of $\mathbf v$. Therefore, for any velocity displacement $\Delta\mathbf{v}$, 
\begin{equation}
\Delta f_m = f_m(\mathbf{v}+\Delta\mathbf{v})-f_m(\mathbf{v})=\mathbf{d}_m^{T}\Delta\mathbf{v}.
\end{equation}
Define the centered covariance of the $m$-th station field over the velocity plane as
\begin{align}\label{r_cm}
r_{c,m}(\Delta\mathbf v)
\triangleq
r_c\left(\mathbf d_m^{T}\Delta\mathbf v\right).
\end{align}
Substituting~\eqref{r_cm} into~\eqref{eq:corr_expansion} gives
\begin{equation}\label{G}
r_{c,m}(\Delta\mathbf{v})
=
1-\frac{1}{2}\lambda_{\mathrm{eff}}
\Delta\mathbf{v}^{T}
\mathbf{d}_m\mathbf{d}_m^{T}
\Delta\mathbf{v}
+o(\|\Delta\mathbf{v}\|^2).
\end{equation}
According to the normalization in~\eqref{length}, the second-order covariance loss associated with a local metric contribution $\mathbf{G}_m$ has the form
\begin{align}\label{1-r_cm}
1-r_{c,m}(\Delta\mathbf{v})
=
\frac{1}{2}
\Delta\mathbf{v}^{T}
\mathbf{G}_m
\Delta\mathbf{v}
+
o\!\left(\|\Delta\mathbf{v}\|_2^2\right).
\end{align}
Comparing~\eqref{1-r_cm} with~\eqref{G} gives
\begin{equation}
\mathbf{G}_m=
\lambda_{\mathrm{eff}}
\mathbf{d}_m\mathbf{d}_m^{T}.
\end{equation}
Since the aggregate second-order covariance loss of the multi-station scan field is the sum of the station-wise losses, 
the overall field-induced metric is
\begin{equation}
\mathbf{G}=\sum_{m=1}^M\mathbf{G}_m=
\sum_{m=1}^M\lambda_{\mathrm{eff}}
\mathbf{d}_m\mathbf{d}_m^{T}.
\label{eq:aggregate_metric}
\end{equation}
Because both $\lambda_{\mathrm{eff}}$ and $\{\mathbf{d}_m\}$ are
independent of the absolute velocity $\mathbf{v}$, the metric in \eqref{eq:aggregate_metric} is
constant over the search region. Furthermore,
\begin{equation}
\mathbf{G} \succ \mathbf{0}
\quad \Longleftrightarrow \quad
\operatorname{span}\{\mathbf{d}_1,\ldots,\mathbf{d}_M\}
=
\mathbb{R}^2.
\end{equation}

Thus, positive definiteness of the two-dimensional field-induced metric requires the station Doppler-sensitivity vectors to span the velocity plane. Physically, this means that the station configuration must provide sensitivity to two noncollinear velocity directions so that both velocity components are locally identifiable.

The centered covariance $r_{c,m}(\Delta\mathbf{v})$ depends only on the velocity displacement $\Delta\mathbf v$, rather than on the absolute velocity $\mathbf v$. Hence, the local correlation structure is translation invariant over the velocity search region. Since the station fields are Gaussian, the joint local distribution of the field and its derivatives is determined by the covariance and its derivatives at zero displacement, and is therefore independent of $\mathbf{v}$. Consequently, the joint local distribution of $T(\mathbf{v})$, $\nabla T(\mathbf{v})$, and $\nabla^2 T(\mathbf{v})$ is also independent of the absolute velocity location. Together with the constancy of $\mathbf{G}$, this establishes the spatial homogeneity of the metric-normalized local Kac–Rice densities required in Lemma~\ref{lem:EED_approx}. The dependence of the one-dimensional Rice density on the metric-unit tangent direction is examined in Section~\ref{sec:ec_densities}.

With the metric $\mathbf G$ now given explicitly by~\eqref{eq:aggregate_metric}, we evaluate the LKCs in~\eqref{eq:L0_def}--\eqref{eq:L2_def} for the connected axis-aligned rectangular velocity search region. Let $W_x$ and $W_y$ denote its side lengths along the $v_x$- and $v_y$-axes, respectively. Substituting the constant metric into the boundary-length and area integrals and using $\chi(\mathcal V)=1$ gives

\begin{align}
L_0(\mathcal{V}) &= 1,\label{L0_value}\\
L_1(\mathcal{V}) &= W_x\sqrt{G_{11}}+W_y\sqrt{G_{22}},\label{L1_value}\\
L_2(\mathcal{V}) &= W_x W_y\sqrt{\det(\mathbf{G})}.\label{L2_value}
\end{align}
The rectangle is contractible, giving $L_0=1$. Its pairs of coordinate-aligned edges have metric lengths $2W_x\sqrt{G_{11}}$ and $2W_y\sqrt{G_{22}}$, while the constant area factor is $\sqrt{\det\mathbf G}$.

Equations \eqref{L0_value}--\eqref{L2_value} provide the geometric quantities required by the EEC approximation. The remaining quantities are the first- and second-order EC densities, which are considered next for the two-station and multi-station cases.

\subsection{EC Densities for Velocity-Plane Scan Field}
\label{sec:ec_densities}

With $\rho_0(\tau)$ available from the pointwise null distribution in Subsection V-A, completing the EEC approximation in Lemma~\ref{lem:EED_approx} requires the boundary and interior densities $\rho_1(\tau)$ and $\rho_2(\tau)$. These depend on the local field derivatives after normalization by the metric in Subsection V-B. We first obtain closed-form densities for a separable two-station geometry, then derive constrained forms for larger networks and calibrate their coefficients using moderate-threshold EEC measurements.

We begin with the two-station geometry used in Section~\ref{sec:eec_simulation}. In metric-normalized coordinates aligned with the station sensitivity directions, the field separates as $T(\xi_1,\xi_2)=A(\xi_1)+B(\xi_2)$, where $A$ and $B$ are independent unit-mean exponential processes. We consider a rectangle aligned with these coordinates, as in Section~\ref{sec:eec_simulation}. The pointwise result in~\eqref{eq:rho0_general} gives
\begin{equation}
\rho_0(\tau)=e^{-\tau}(1+\tau).
\end{equation}

For the boundary term, $\partial T/\partial\xi_1=A'$. Gaussian conditioning gives $A'\mid A=s\sim\mathcal N(0,2s)$, so $\mathbb E[(A')_+\mid A=s]=\sqrt{s/\pi}$. Since $p_T(\tau)=\tau e^{-\tau}$ and $A\mid T=\tau$ is uniform on $[0,\tau]$, the first-order Rice formula in Lemma~\ref{lem:EED_approx} gives
\begin{equation}
\begin{aligned}
\rho_1(\tau)&=\frac{p_T(\tau)}{\tau}\int_0^\tau\sqrt{\frac{s}{\pi}}\,ds\\
&=\frac{2}{3\sqrt{\pi}}e^{-\tau}\tau^{3/2}.
\end{aligned}
\label{eq:review_two_station_boundary}
\end{equation}
\noindent

For the interior term, $\mathbb E[-A''\mid A=s,A'=0]=2s-1$ yields the signed one-dimensional critical-height density $q(s)=e^{-s}(2s-1)/(2\sqrt{\pi s})$. Independence and separability imply that contributions at heights $s$ and $u-s$ multiply. Integrating over these heights and then over total heights above $\tau$ gives
\begin{equation}
\begin{aligned}
\rho_2(\tau)&=\int_\tau^\infty\!\int_0^u q(s)q(u-s)\,ds\,du\\
&=\frac{1}{8}e^{-\tau}\tau(\tau-2).
\end{aligned}
\label{eq:review_two_station_convolution}
\end{equation}

We next consider networks with $M\geq3$. The pointwise statistic remains Gamma distributed, and hence the zeroth-order EC density is available in closed form as in~\eqref{eq:rho0_general},
but direct evaluation of the higher-order EC densities is more involved because it requires the joint local statistics of $T(\mathbf{v})$, its gradient, and its Hessian. Direct Monte Carlo estimation of the desired rare-event false-alarm probability is also inefficient at high thresholds. We therefore identify a small number of threshold-independent EC-density parameters from subdomain EEC measurements at moderate thresholds and subsequently use their theoretically constrained tail forms for high-threshold extrapolation. Let
\begin{equation}
U_m(\mathbf{v})
\triangleq
|\eta_m(\mathbf{v})|^2,
\qquad
T(\mathbf{v})
=
\sum_{m=1}^{M}
U_m(\mathbf{v}).
\end{equation}
Conditioned on $T=u$, the station-wise component energies can be represented as
\begin{equation}
U_m = uP_m,
\end{equation}
where the change of variables $U_m=uP_m$ has Jacobian $u^{M-1}$, so the independent exponential densities factor into a Gamma density for $u$ and a uniform density on the proportion simplex:
\begin{equation}
(P_1,\ldots,P_M)
\sim
\operatorname{Dirichlet}(1,\ldots,1),
\end{equation}
and the distribution of the proportions $\{P_m\}$ is independent of $u$. This representation separates the total-energy scale $u$ from the relative energy allocation among the stations in the Kac--Rice conditional moments.

For the first-order EC density, the pointwise Gamma density contributes the factor
\begin{equation}
p_T(u)
=
\frac{e^{-u}u^{M-1}}{\Gamma(M)},
\end{equation}
while the conditional standard deviation of a metric-normalized directional derivative scales as $u^{1/2}$. Conditional on the component energies, this derivative is zero-mean Gaussian, so its mean positive part also scales as $u^{1/2}$. Under the tangent-direction-independence condition in Lemma~\ref{lem:EED_approx}, the first-order EC density has the form
\begin{equation}
\rho_1(\tau)
=
c_{1,M} e^{-\tau}\tau^{M-\frac{1}{2}},
\end{equation}
where $c_{1,M}$ is independent of $\tau$ and collects the remaining local derivative statistics. In general, this coefficient may depend on the normalized station geometry, and this dependence is suppressed in the notation.

For the second-order density, define the level-specific Kac--Rice integrand

\begin{equation}
\kappa_2(u)
\triangleq
p_{T,\nabla_{\xi}T}(u,\mathbf{0})
\,
\mathbb{E}
\left[
\det\!\left(-\nabla_{\xi}^{2}T\right)
\,\middle|\,
T=u,\,
\nabla_{\xi}T=\mathbf{0}
\right].
\end{equation}
To obtain its threshold dependence, fix the energy proportions $\mathbf P$ and define $\mathbf S(\mathbf P)=\mathbf G^{-1/2}(\sum_m P_m\mathbf G_m)\mathbf G^{-1/2}$. The conditional gradient is Gaussian, with
\begin{equation}
\begin{aligned}
\nabla_\xi T\mid(T=u,\mathbf P)&\sim\mathcal N(\mathbf0,2u\mathbf S(\mathbf P)),\\
p_{\nabla_\xi T\mid T,\mathbf P}(\mathbf0\mid u,\mathbf P)
&=\frac{1}{4\pi u\sqrt{\det\mathbf S(\mathbf P)}}.
\end{aligned}
\label{eq:review_gradient_scaling}
\end{equation}
\noindent After division by $\sqrt u$, the zero-gradient constraint is independent of $u$. The even centered covariance makes the Gaussian first derivatives independent of the field values and second derivatives. Gaussian regression then gives, conditional on $T=u$, $\mathbf P$, and $\nabla_\xi T=\mathbf0$,
\begin{equation}
\nabla_\xi^2T\overset{d}{=}\mathbf H_0+\sqrt u\,\mathbf H_1+u\mathbf H_2,
\label{eq:review_hessian_scaling}
\end{equation}
\noindent where $\mathbf H_2=-2\mathbf S(\mathbf P)$, $\mathbf H_1$ is zero-mean and independent of $\mathbf H_0$, and their conditional laws do not depend on $u$. Thus, the odd powers of $\sqrt u$ vanish in the expected determinant, leaving $\sum_{j=0}^2 b_j(\mathbf P)u^j$. Multiplying by the zero-gradient density and $p_T(u)$, then averaging their product over the Dirichlet proportions, gives
\begin{equation}
\kappa_2(u)
=
e^{-u}
\sum_{j=0}^{2}
\theta_{j,M}u^{M-2+j},
\end{equation}
where the coefficients $\theta_{j,M}$ are independent of $u$. Since
\begin{equation}
\rho_2(\tau)
=
\int_{\tau}^{\infty}
\kappa_2(u)\,du,
\end{equation}
we obtain the constrained incomplete Gamma representation

\begin{equation}
\rho_2(\tau)
=
\sum_{j=0}^{2}
\theta_{j,M}
\Gamma(M-1+j,\tau),
\end{equation}
where 
\begin{equation}
\Gamma(s,\tau)
\triangleq
\int_{\tau}^{\infty}
u^{s-1}e^{-u}\,du
\end{equation}
is the upper incomplete Gamma function. For integer $s$,
\begin{equation}
\Gamma(s,\tau)
=
(s-1)!
e^{-\tau}
\sum_{k=0}^{s-1}
\frac{\tau^k}{k!},
\end{equation}
and hence the above representation is equivalent to an exponential-polynomial form. However,
unlike an unrestricted degree-$M$ polynomial, the incomplete-gamma representation constrains the
polynomial coefficients to the three-dimensional subspace implied by the Kac--Rice integrand.

With the threshold dependence specified, we now determine the remaining coefficients from moderate-threshold EEC measurements. We construct $\{\mathcal{V}_d\}_{d=1}^{D}$ as centered, axis-aligned subrectangles of $\mathcal V$. Their LKCs follow from~\eqref{L0_value}--\eqref{L2_value}. For each calibration threshold $\tau_j$ and Monte Carlo realization $r$, we compute

\begin{equation}
\chi\!\left(A_{\tau_j}^{(r)} \cap \mathcal{V}_d\right).
\end{equation}
The corresponding subdomain EEC is estimated as
\begin{equation}
\widehat{\mu}_d(\tau_j)
=
\frac{1}{R}
\sum_{r=1}^{R}
\chi\!\left(A_{\tau_j}^{(r)} \cap \mathcal{V}_d\right),
\end{equation}
where $R$ is the number of Monte Carlo simulations. The calibration interval is selected such that the threshold is sufficiently high for the excursion set to
be dominated by isolated components, while excursions remain frequent enough for reliable Monte
Carlo estimation. Thus, the Monte Carlo simulation is performed only in a moderate-threshold
regime rather than directly in the desired rare-event regime.

Because $\rho_0(\tau)$ is known analytically, define
\begin{equation}
\widetilde{\mu}_d(\tau_j)
\triangleq
\widehat{\mu}_d(\tau_j)
-
L_0(\mathcal{V}_d)\rho_0(\tau_j).
\end{equation}
Substituting the constrained expressions for $\rho_1$ and $\rho_2$ into the EEC decomposition gives
\begin{align}
\widetilde{\mu}_d(\tau_j)
&\approx
L_1(\mathcal{V}_d)c_{1,M}e^{-\tau_j}\tau_j^{M-\frac{1}{2}}
\nonumber\\
&\quad
+
L_2(\mathcal{V}_d)
\sum_{q=0}^{2}
\theta_{q,M}\Gamma(M-1+q,\tau_j).
\end{align}

The four threshold-independent coefficients are jointly estimated over all calibration subdomains and thresholds:

\begin{multline}
\left[
\widehat{c}_{1,M},
\widehat{\theta}_{0,M},
\widehat{\theta}_{1,M},
\widehat{\theta}_{2,M}
\right]
\\=
\arg\min_{c,\boldsymbol{\theta}}
\sum_{d,j}
w_{d,j}
\Bigg[
\widetilde{\mu}_d(\tau_j)
-
L_1(\mathcal{V}_d)c
e^{-\tau_j}\tau_j^{M-\frac{1}{2}}
\nonumber\\
\\ \qquad\qquad\qquad
-
L_2(\mathcal{V}_d)
\sum_{q=0}^{2}
\theta_{q,M}
\Gamma(M-1+q,\tau_j)
\Bigg]^2,
\end{multline}
where $\boldsymbol{\theta} = [\theta_{0,M},\theta_{1,M},\theta_{2,M}]^T$ and $w_{d,j}$ may be selected according to the Monte Carlo variance of the subdomain EEC estimate, with uniform weights giving ordinary least squares (LS). In the numerical implementation, the complete search region is excluded from coefficient identification and is retained as a domain-level validation region.

For the axis-aligned rectangular domains used in the simulations, an auxiliary direction-resolved regression separates the two coordinate-aligned boundary contributions. A common coefficient $c_{1,M}$ is used only when their estimates agree within the Monte Carlo uncertainty. This check concerns these boundary orientations and does not establish isotropy for arbitrary directions.

After calibration, the EC densities are evaluated at arbitrary thresholds as
\begin{equation}
\widehat{\rho}_1(\tau)
=
\widehat{c}_{1,M}
e^{-\tau}
\tau^{M-\frac{1}{2}},
\end{equation}
and
\begin{equation}
\widehat{\rho}_2(\tau)
=
\sum_{q=0}^{2}
\widehat{\theta}_{q,M}
\Gamma(M-1+q,\tau).
\end{equation}
For the full velocity search region,
\begin{equation}
\widehat{\mu}(\tau)
=
L_0(\mathcal{V})\rho_0(\tau)
+
L_1(\mathcal{V})\widehat{\rho}_1(\tau)
+
L_2(\mathcal{V})\widehat{\rho}_2(\tau),
\end{equation}
and the false-alarm probability is approximated as
\begin{equation}
\widehat{P}_{\mathrm{FA}}(\tau)
\approx
1-\exp\left[-\widehat{\mu}(\tau)\right].
\end{equation}
The CFAR threshold for a prescribed false-alarm level $\alpha$ is then obtained by solving
\begin{equation}
\widehat{P}_{\mathrm{FA}}(\tau_{\mathrm{CFAR}})
=
\alpha.
\end{equation}

This procedure uses moderate-threshold Monte Carlo simulations only to identify a small number of local EC-density parameters. The model-specific constrained incomplete Gamma structure then provides the extrapolation to the high-threshold rare-event regime, without requiring direct Monte Carlo estimation at the target false-alarm probability.

\section{Simulation Results}
\label{sec:simulations}

In this section, we use Monte Carlo (MC) simulations to evaluate the vanishing quantization loss, the detection gain of geometry-consistent velocity search, and the accuracy of the EEC-based false-alarm calibration.

\subsection{Vanishing Quantization Loss}
\label{sec:vanishing_quantization_loss}
\label{sec:quantization_grid}

We first examine the vanishing quantization loss predicted in Section~\ref{sec:detector}. A transmitter at $(0,0)$~m and three receivers at $(0,0)$, $(10000,0)$, and $(0,5000)$~m observe a target at $(5000,5000)$~m moving at $(15,-9)$~m/s. The carrier frequency is $24$~GHz, with $30$ pulses per CPI and a pulse-repetition frequency (PRF) of $10$~kHz. Each $2$~$\mu$s pulse contains $2400$ samples and the single-frequency pulses are stepped from $300$ to $600$~MHz.

Each receiver observes $100$ compound-Gaussian scatterers. Textures follow a Gamma law with shape $0.8$ and scale $1$, speckle coefficients follow $\mathcal{CN}(0,1)$, and Dopplers are uniform on $[-800,800]$~Hz. Texture, speckle coefficients, delays, and Dopplers are fixed within each CPI and independently redrawn across scatterers, receivers, and Monte Carlo CPIs. Sample-grid delays span approximately $\pm1$~$\mu$s about the reference gate, retaining only the overlapping pulse portions. Echoes vary across pulses through Doppler phase and the stepped waveform. Receiver noise is independent across samples and pulses. Scenario~II reduces clutter power by $15$~dB relative to Scenario~I.

Conditional on texture, delays, and Dopplers, Gaussian speckle supplies the clutter randomness. Under the weak-input approximation, the high-resolution and one-bit ADC branches share the same target and clutter terms after rescaling, without relying on the Gamma form. One representative texture law therefore suffices to illustrate vanishing loss under post-filtering clutter dominance. Texture statistics can still affect finite-SNR performance and how often these conditions hold.

The two branches use the same realizations and unwhitened velocity matched filter, with both velocity components searched over $[-22.097,22.097]$~m/s on a common $65\times65$ grid. Each branch is calibrated separately at $P_{\rm FA}=10^{-2}$ using $10^4$ null realizations. Results use $5000$ independent null realizations and $4000$ signal-present realizations per SNR. The achieved false-alarm rates range from $0.0096$ to $0.0104$. Refining the grid to $129\times129$ and recalibrating the thresholds changed $P_{\rm D}$ by at most $1.2$ percentage points at the two plotted SNRs. Input SNR denotes the target-to-receiver-noise power ratio. The required-SNR gap at $P_{\rm D}=0.8$ is obtained by linear interpolation of the $1$~dB SNR sweep. Paired-bootstrap confidence intervals include calibration and detection uncertainty but exclude SNR interpolation error.

\begin{table}[!ht]
\centering
\caption{Interference levels and quantization loss at $P_{\rm FA}=10^{-2}$ and $P_{\rm D}=0.8$.}
\label{tab:vi_a_diagnostics}
\begin{tabular}{lcc}
\toprule
Quantity & Scenario I & Scenario II\\
\midrule
Input CNR (dB) & $-15.24$ & $-30.24$\\
Matched-filter CNR (dB) & $17.47$ & $2.47$\\
True-direction CNR (dB) & $17.53$ & $2.38$\\
Required-SNR gap (dB) & $-0.01$ & $0.40$\\
95\% confidence interval (dB) & $[-0.04,0.07]$ & $[0.29,0.48]$\\
\bottomrule
\end{tabular}
\end{table}

Table~\ref{tab:vi_a_diagnostics} reports the clutter-to-noise ratios (CNRs) before and after matched filtering. In Scenario~I, the input CNR is $-15.24$~dB, whereas the matched-filter CNR reaches $17.47$~dB. Thus, weak clutter at the ADC input coexists with clutter-dominated interference after matched filtering. The CNR along the true target direction is $17.53$~dB, confirming that this dominance also holds in the relevant detection direction.

\begin{figure}[!t]
\centering
\includegraphics[width=\columnwidth]{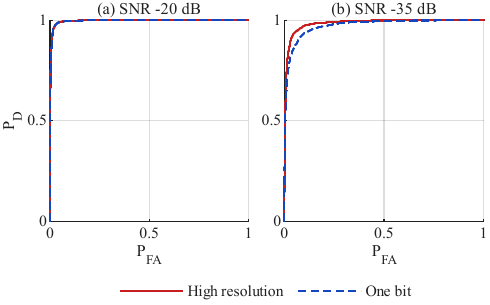}
\caption{High-resolution and one-bit ROC curves. (a) Scenario~I. (b) Scenario~II.}
\label{fig:vi_a_roc}
\end{figure}

\begin{figure}[!t]
\centering
\includegraphics[width=\columnwidth]{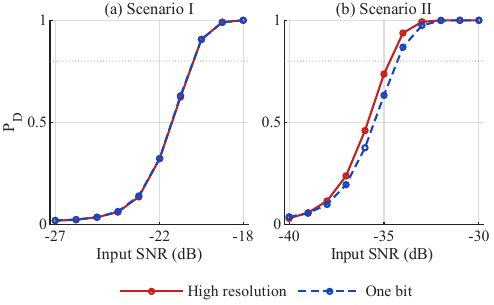}
\caption{Detection probability versus input SNR at $P_{\rm FA}=10^{-2}$. (a) Scenario~I. (b) Scenario~II. The horizontal line marks $P_{\rm D}=0.8$.}
\label{fig:vi_a_pd}
\end{figure}

Figure~\ref{fig:vi_a_roc}(a) shows nearly coincident high-resolution and one-bit ROC curves in Scenario~I. The detection curves in Fig.~\ref{fig:vi_a_pd}(a) likewise overlap: at $P_{\rm D}=0.8$, the required-SNR gap is $-0.01$~dB, with a 95\% confidence interval of $[-0.04,0.07]$~dB. This result supports vanishing quantization loss within the resolution of the experiment. In Scenario~II, the matched-filter CNR decreases to $2.47$~dB, and the curves separate, as shown in Figs.~\ref{fig:vi_a_roc}(b) and~\ref{fig:vi_a_pd}(b). The resulting gap is $0.40$~dB, still well below the classical low-SNR noise-limited loss of approximately $1.96$~dB~\cite{xiao2022onebit}. Together, these results support the central prediction: one-bit detection approaches high-resolution performance as post-filtering clutter increasingly dominates the additive noise.

\subsection{Geometry-Consistent Velocity Search}
\label{sec:velocity_search_gain}
We next assess the role of joint velocity search in the effective one-bit slow-time model of Section~\ref{sec:velocity_search}, using its normalized white-interference specialization with independent $\mathcal{CN}(\mathbf{0},\mathbf{I}_{30})$ disturbances. The three-receiver geometry, carrier frequency, PRF, and search region are unchanged, while a second geometry adds a receiver at $(10000,8000)$~m. Target velocity components are drawn independently and uniformly on $[-18,18]$~m/s. Network SNR is the sum of the station matched-filter SNRs, expressed in decibels, with equal signal energy across receivers. It is held fixed when the receiver count changes.

The Doppler-estimation and LS velocity-fitting stage of~\cite{xiang2023liquid} motivates the estimate-then-fuse (ETF) baselines below, which we construct using the same slow-time data and station scores as the velocity-plane (VP) detector. ETF-LS selects each station's strongest Doppler peak, fuses the estimates by box-constrained LS, and rescores the fitted velocity. ETF3-LS retains the three strongest local peaks per station and evaluates all fused combinations. Independent Doppler fusion (IDF) sums the independently maximized station scores, without enforcing a common velocity. A known-velocity detector provides a reference without search uncertainty. VP uses a $65\times65$ grid with local refinement and retains the ETF3-LS candidates. All methods use the marginal Doppler ranges of the same velocity region. Each method is calibrated at $P_{\rm FA}=10^{-2}$ using $20000$ null trials, then evaluated on $20000$ independent null trials and $5000$ signal-present trials per SNR. Achieved false-alarm rates range from $0.0087$ to $0.0109$.

\begin{figure}[!t]
\centering
\includegraphics[width=\columnwidth]{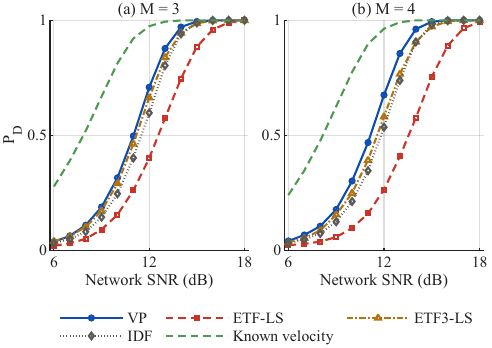}
\caption{Detection probability at $P_{\rm FA}=10^{-2}$ with fixed total network SNR. (a) Three receivers. (b) Four receivers.}
\label{fig:vi_b_detection}
\end{figure}

Figure~\ref{fig:vi_b_detection} shows that, at $P_{\rm D}=0.8$, VP reduces the required SNR relative to ETF-LS by $1.85$ and $2.64$~dB for three and four receivers, respectively. Retaining multiple station peaks narrows the gap, but VP still gains $0.24$ and $0.55$~dB over ETF3-LS. The corresponding gains over IDF are $0.44$ and $0.68$~dB. These results support the joint velocity search used in Section~\ref{sec:velocity_search}: combining station scores before selecting a common velocity remains beneficial under the effective one-bit model, while retaining multiple local peaks recovers part of this gain.

\subsection{EEC-Based False-Alarm Calibration}
\label{sec:eec_simulation}
We use independent $\mathcal{CN}(0,1)$ slow-time samples, with $N$, PRF, and wavelength as in Section~\ref{sec:vanishing_quantization_loss}. The transmitter is located at $(-5,-5)$~km and the range gate at the origin. Both velocity components span $[-11.05,11.05]$~m/s on a $257\times257$ grid. We denote the fraction of null realizations whose maximum score exceeds $\tau$ by $p_{\rm MC}(\tau)$ and the sample mean of the excursion-set Euler characteristic by $\widehat\mu_{\rm MC}(\tau)$.

\emph{Two-station case:} Receivers at $(-5,5)$ and $(5,-5)$~km yield a separable velocity field. The closed-form EEC based on the axis-aligned densities in~\eqref{eq:review_two_station_boundary} and~\eqref{eq:review_two_station_convolution} is evaluated without fitting and validated using $2\times10^7$ independent realizations. Paired refinement to $513\times513$ shifts the tested MC reference thresholds by less than $0.005$.

\begin{figure}[!t]
\centering
\includegraphics[width=\columnwidth]{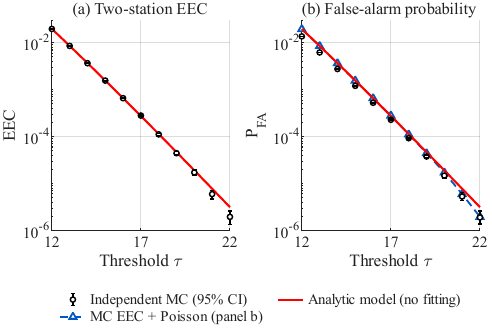}
\caption{Two-station validation. (a) EEC. (b) $P_{\rm FA}$ with the MC EEC--Poisson diagnostic. Error bars indicate $95\%$ confidence intervals.}
\label{fig:vi_c_m2}
\end{figure}

Figure~\ref{fig:vi_c_m2}(a) compares the analytic and MC EECs. At the six tested targets from $\alpha=10^{-2}$ to $3\times10^{-5}$, the prediction lies within the pointwise MC $95\%$ intervals. It exceeds these intervals at $10^{-5}$ and at the larger integer thresholds $\tau=\{21,22\}$. In Fig.~\ref{fig:vi_c_m2}(b), $1-\exp[-\widehat\mu_{\rm MC}]$ overpredicts $P_{\rm FA}$ by $38.2\%$ at $\alpha=10^{-2}$ and $12.7\%$ at $10^{-5}$, relative to direct MC $P_{\rm FA}$. The latter discrepancy has a paired $95\%$ interval of $[5.2,20.1]\%$. At $\alpha=10^{-5}$, $\tau_{\rm EEC}=20.718$ produces $150$ false alarms, giving $p_{\rm MC}=7.50\times10^{-6}$ ($95\%$ interval: $[6.39,8.80]\times10^{-6}$). The conversion discrepancy is smaller at the selected higher operating thresholds, although the achieved $P_{\rm FA}$ remains conservative and its deviation from the nominal level is not monotone.

\begin{figure}[!ht]
\centering
\includegraphics[width=\columnwidth]{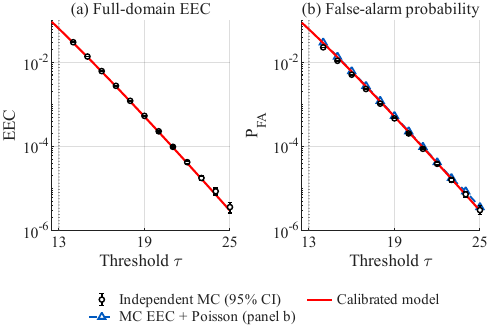}
\caption{Three-station validation. (a) EEC. (b) $P_{\rm FA}$ with the MC EEC--Poisson diagnostic. The dotted line marks the maximum fitting threshold. Error bars indicate $95\%$ confidence intervals.}
\label{fig:vi_c_calibration}
\end{figure}

\emph{Three-station case:} A third receiver is added at the transmitter. Boundary densities agree along the two rectangle directions by symmetry.

Four threshold-independent coefficients are fitted using $10^5$ realizations, five proper subrectangles, and seven thresholds from $8.5$ to $13$. The full domain and all higher thresholds are excluded from fitting. The frozen model is validated using $2\times10^7$ new independent realizations at thresholds up to $25$.

Figure~\ref{fig:vi_c_calibration} compares EEC and $P_{\rm FA}$ predictions with independent MC. The estimated $1-\exp[-\widehat\mu_{\rm MC}]$ overpredicts $P_{\rm FA}$ by $24.0\%$ at $\alpha=10^{-2}$, $11.5\%$ at $10^{-4}$, and $8.9\%$ at $3\times10^{-5}$, relative to direct MC. At $10^{-5}$, the discrepancy is $15.6\%$, with a paired $95\%$ interval of $[7.4,23.9]\%$. Thus, the conversion improves overall without decreasing at every tested level. At $\alpha=10^{-5}$, the fitted prediction differs from this diagnostic by $-11.5\%$ of direct MC $P_{\rm FA}$, so overall agreement still partly reflects error cancellation. In a paired grid check, refinement to $513\times513$ shifts the MC reference thresholds by less than $0.008$ for the tested $\alpha\le10^{-3}$. At $\alpha=10^{-5}$, the relative EEC-to-$P_{\rm FA}$ discrepancy changes from $15.6\%$ to $12.5\%$.

Table~\ref{tab:vi_c_thresholds} reports thresholds and achieved $P_{\rm FA}$ values. The MC reference threshold is the ascending order statistic with rank $\lceil(n+1)(1-\alpha)\rceil$, where $n=2\times10^7$, from the same validation sample. From $10^{-2}$ to $10^{-5}$, the EEC--MC threshold gap decreases from $0.247$ to $0.040$, and the $P_{\rm FA}$ deficit from $16.9\%$ to $4.0\%$. At $10^{-5}$, $192$ false alarms yield $p_{\rm MC}=9.60\times10^{-6}$. These results support low-$P_{\rm FA}$ extrapolation with residual conversion error and sampling uncertainty.

\begin{table}[!ht]
\centering
\caption{Three-station EEC thresholds and MC references. $\Delta\tau=\tau_{\rm EEC}-\tau_{\rm MC}$; $\delta_\alpha=100(p_{\rm MC}/\alpha-1)$. $P_{\rm FA}$ estimates and Wilson $95\%$ intervals at $\tau_{\rm EEC}$ are normalized by $\alpha$.}
\label{tab:vi_c_thresholds}
\begingroup
\footnotesize
\setlength{\tabcolsep}{3pt}
\renewcommand{\arraystretch}{1.15}
\begin{tabular}{ccccccc}
\toprule
$\alpha$ & $\tau_{\rm EEC}$ & $\tau_{\rm MC}$ & $\Delta\tau$ & $p_{\rm MC}/\alpha$ & $95\%\ \mathrm{CI}$ & $\delta_\alpha\ (\%)$\\
\midrule
$10^{-2}$ & $15.359$ & $15.112$ & $0.247$ & $0.8307$ & $[0.827,0.835]$ & $-16.9$\\
$10^{-3}$ & $18.195$ & $18.046$ & $0.148$ & $0.8917$ & $[0.879,0.905]$ & $-10.8$\\
$10^{-4}$ & $20.932$ & $20.851$ & $0.081$ & $0.9345$ & $[0.893,0.978]$ & $-6.6$\\
$10^{-5}$ & $23.605$ & $23.565$ & $0.040$ & $0.9600$ & $[0.833,1.106]$ & $-4.0$\\
\bottomrule
\end{tabular}
\endgroup
\end{table}

\section{Conclusion}
\label{sec:conclusion}

The central finding of this work is that one-bit quantization loss can vanish in clutter-limited radar detection. When target and clutter echoes are weak at the ADC input, one-bit sampling approximately preserves their relative strength. If clutter subsequently dominates the effective additive noise in the detection subspace, one-bit and high-resolution detection of a known signature approach the same performance under the Gaussian-equivalent model with known covariances. Simulations show negligible loss for the velocity-scan detector in the tested clutter-dominated setting, together with detection gains over estimate-then-fuse baselines. EEC-based threshold calibration combines analytic expressions for a separable two-station geometry with coefficient fitting for larger networks. Independent validation supports extrapolation to nominal false-alarm probabilities of $10^{-5}$ in the tested Gaussian-equivalent fields, while revealing residual EEC-to-$P_{\rm FA}$ conversion error. These results identify a regime in which one-bit sampling can reduce data volume with negligible detection loss.

\section*{Acknowledgment}
ChatGPT and Codex assisted with manuscript preparation and simulation-code development in Section VI.

\bibliographystyle{IEEEtran}
\bibliography{verified_references}

\end{document}